\documentclass[aps,pra,reprint,showpacs,showkeys]{revtex4-2}
\usepackage{amsmath, amssymb, amsthm}
\usepackage{graphicx}
\usepackage{hyperref}
\usepackage[utf8]{inputenc}
\usepackage[english]{babel}
\usepackage{tikz}
\usetikzlibrary{shapes.geometric, arrows, fit, backgrounds}
\usepackage{booktabs}

\begin{document}
\title{Predicting the von Neumann Entanglement Entropy Using a Graph Neural Network}
\author{Anas Saleh}
\email{anas-saleh@uiowa.edu}
\affiliation{Department of Physics and Astronomy, University of Iowa, Iowa City, Iowa, USA}

\date{\today}

\begin{abstract}
Calculating the von Neumann entanglement entropy from experimental data is challenging due to its dependence on the complete wavefunction, forcing reliance on approximations such as classical mutual information (MI). We propose a machine learning approach using a graph neural network to predict the von Neumann entropy directly from experimentally accessible bitstrings. We test this approach on a Rydberg ladder system and achieve a mean absolute error of $3.6 \times 10^{-3}$ when evaluating within the training range on a dataset with entropy values ranging from 0 to 1.9. The model achieves a mean absolute percentage error of 1.44\% and outperforms MI-based bounds. When tested beyond the training range, the model maintains reasonable accuracy. Furthermore, we demonstrate that fine-tuning the model with small datasets significantly improves performance on data outside the original training range.
\end{abstract}
\maketitle

\section{Introduction}

Quantum entanglement is a cornerstone of quantum mechanics and a widely studied subject\cite{meurice2024experimental,sun2024genuine,huang2025integrated,barr2024quantum}, and the von Neumann entanglement entropy serves as a key measure of entanglement with broad applications across physics, it's defined as
\begin{align}
S^{vN}_A=-\text{Tr}(\rho_A \ln(\rho_A)),	
\end{align}     
where $\rho_A = \text{Tr}_B \rho_{AB}$ is the reduced density matrix of subsystem A. 

However, experimental measurement of $S^{\text{vN}}_A$ faces a fundamental challenge: it requires complete knowledge of the quantum wavefunction, which is not directly accessible through measurements. Traditional approaches rely on full quantum state tomography, which scales exponentially with system size and becomes prohibitively expensive for many-body systems. This scaling bottleneck severely limits our ability to probe entanglement in the large-scale quantum systems where many-body phenomena emerge.

Recent work has begun addressing this challenge by developing bounds and approximations for entanglement entropy. Notably, a study of Rydberg ladder systems demonstrated that classical mutual information (MI) derived from measurement bitstrings can provide a lower bound for $S^{\text{vN}}_A$~\cite{meurice2024experimental}. While this approach represents significant progress, MI-based bounds are often loose and may not capture the full entanglement structure.

In this work, we propose a fundamentally different approach: using graph neural networks (GNNs) to predict $S^{\text{vN}}_A$ directly from the same experimental bitstring data used for MI calculations. Our method bypasses the need for full wavefunction reconstruction by leveraging the correlation structure embedded in measurement outcomes. 

GNNs provide a natural framework for this task because their graph structure naturally mirrors quantum systems: nodes represent individual particles while edges encode the correlations and interactions that give rise to entanglement. Unlike tomography, which requires exponentially many measurements, our approach can predict entanglement entropy from polynomial-scale experimental data, making it practically feasible for larger quantum systems.

The paper is structured as follows: Section 2 reviews machine learning in quantum physics and GNNs, Section 3 details our methodology (including data generation and model architecture), and Section 4 presents our results and demonstrates how fine-tuning the model improves performance after initial training.

\section{Background}

\subsection{Machine Learning for Quantum Systems}

Machine learning (ML) attracted interest from all fields of study and showed its ability to tackle many complex problems \cite{shi2023one, shi2024unified, zhou2025resource, zhang2023planet}. Quantum physics has emerged as one of the heavily affected fields and it was shown that ML can offering novel solutions to problems once deemed intractable due to the inherent complexity of quantum systems. Quantum mechanics, governed by principles such as superposition, entanglement, and exponential state-space scaling, poses unique challenges for both theoretical modeling and experimental data analysis. Traditional computational methods, such as exact diagonalization or quantum Monte Carlo, often struggle with the "curse of dimensionality" in many-body systems or the noise susceptibility of quantum hardware. Machine learning, with its capacity to approximate high-dimensional functions, extract patterns from sparse data, and optimize complex systems, has become an indispensable tool in advancing quantum research\cite{carleo2017solving}.

In quantum many-body physics, neural networks revolutionized the representation of wavefunctions. For instance, neural quantum states (NQS) leverage deep learning architectures to approximate ground and excited states of correlated quantum systems, bypassing the limitations of conventional variational methods. These approaches have enabled precise simulations of spin lattices, fermionic systems, and quantum phase transitions\cite{carleo2019machine,pfau2020ab}.

Quantum computing has also benefited from ML. In \cite{fosel2018reinforcement}, it was shown that neural-network-based reinforcement learning (RL) can discover quantum error correction (QEC) strategies for protecting logical qubits against noise in few-qubit systems.

Meanwhile, generative models like quantum autoencoders assist in quantum error correction, leading to extended lifetimes of logical qubits\cite{locher2023quantum}. These advances are critical for bridging the gap between theoretical quantum advantage and practical implementation.

\subsection{Graph Neural Networks in Physics}
Graph Neural Networks have emerged as a powerful framework for modeling systems with inherent relational or topological structure, such as atomic lattices, particle interaction networks, and various other physical systems. Unlike traditional neural architectures, GNNs operate directly on graph-structured data, leveraging message-passing mechanisms to propagate and aggregate information between nodes and edges\cite{hu2019strategies,shi2020masked,huang2024flow2gnn}. This capability aligns naturally with physics problems where locality, symmetry, and connectivity play defining roles, such as in material science, quantum chemistry, and particle physics \cite{battaglia2018relational}.

Examples of GNN applications in physics include predicting material properties in condensed matter physics, as they are especially suited for such tasks\cite{xie2018crystal}, and top tagging as discussed in\cite{qu2020jet}. In quantum physics applications, GNNs have been successfully applied to particle track reconstruction in high-energy physics and quantum state analysis. They have been implemented in \cite{rieger2024sample} to predict the Rényi entropy, demonstrating their versatility across different domains of physical sciences.

The field has also seen significant developments in quantum machine learning frameworks that exploit graph structures. Recent works have established quantum graph neural networks (QGNNs) as powerful tools for quantum machine learning of graph-structured data, with applications ranging from materials science to high-energy physics\cite{ryu2023quantum,tuysuz2021hybrid}. These approaches leverage the natural graph representation of quantum systems to improve learning performance through specialized quantum neural network architectures.

\section{Methodology}
\subsection{Dataset}
In this work, we studied a Rydberg system described by the following Hamiltonian:
\begin{align}
\begin{split}
H = \frac{\Omega}{2} \sum_{i} \biggl( e^{i\phi} \,\lvert g\rangle_{i}\langle e\rvert_{i} 
+ e^{-i\phi} \,\lvert e\rangle_{i}\langle g\rvert_{i} \biggr)\\
- \Delta(t)\sum_{i} \hat{n}_{i}
+ \sum_{i<j} \frac{C_{6}}{\lvert \vec{r}_{i} - \vec{r}_{j}\rvert^{6}} \,\hat{n}_{i}\hat{n}_{j}.
\end{split}
\end{align}
Where $C_6 = 5.42 \times 10^{-24} \frac{\text{m}^6}{\text{s}} \text{rad}$, $\Omega$ is the Rabi drive amplitude, and $\Delta$ is the detuning.

The atoms were arranged in a two-leg ladder configuration, with the $y$-separation being twice the $x$-separation. This system is based on the Aquila quantum computer developed by QuEra.

To generate our dataset, we randomly sampled specific ranges of the Hamiltonian parameters. But instead of sampling over $\Delta$, $\Omega$, and $a$ (where $a$ is the lattice $x$-spacing), we normalized our Hamiltonian by dividing by $\Omega$. Then, we sampled the dimensionless ratios $\frac{\Delta}{\Omega}$ and $\frac{R_b}{a}$, where $R_b$ is defined as $R_b = \left(\frac{C_6}{\Omega}\right)^{1/6}$. Additionally, we sampled the number of rungs (ladder legs) in the system. The sampling ranges were as follows:
\begin{align}
\begin{split}
\frac{\Delta}{\Omega} &\in [0,6],\\
\frac{R_b}{a} &\in [0.1,5],\\
N_{\text{rungs}} &\in [1,6].
\end{split}
\end{align}
The number of samples taken for $\frac{\Delta}{\Omega}$ and $\frac{R_b}{a}$ were evenly distributed. However, for the number of rungs, we increased the number of samples for larger systems to account for the growing complexity with system size. Specifically, the number of samples for rungs 1 through 6 were 30,000, 60,000, 100,000, 200,000, 350,000, 500,000, respectively. These sampled regions span three distinct ordered phases along with a disordered phase covering a region of great theoretical interest\cite{keesling2019quantum}.

For each sample, we diagonalized the Hamiltonian, computed the ground state, randomly selected a subsystem, and calculated the von Neumann entanglement entropy.

\subsection{Preprocessing}  
Each data point was transformed into a fully connected graph, where nodes represent the Rydberg atoms, and edges represent the interactions between these atoms. We then defined a set of experimentally accessible features for the nodes, edges, and the graph as a whole.  

\subsubsection{Node Features}  
For each node, we assigned the following features:  
\begin{itemize}  
    \item \textbf{Coordinates}: The atoms were placed on a grid with $x$-spacing = 1 and $y$-spacing = 2.  
    \item \textbf{Rydberg State Probability}: The probability of the atom being in the Rydberg state, calculated using the ground state probabilities.  
    \item \textbf{Subsystem Mask}: A binary value indicating whether the node belongs to subsystem `A' or `B'.  
\end{itemize}  

\subsubsection{Edge Features}  
 For each edge, we defined the following features:  
\begin{itemize}  
    \item \textbf{Normalized Distance}: The distance between the connected nodes divided by the square root of the total number of atoms.  
    \item \textbf{Angle with Horizontal Axis}: The angle that the edge makes with the horizontal axis.  
    \item \textbf{Two-Point Correlation}: Defined as $C_{ij} = \langle r_i r_j \rangle - \langle r_i \rangle \langle r_j \rangle$.  
\end{itemize}  

The first two features were chosen to encode the geometric structure of the system, while the third feature captures the correlations between Rydberg atoms.  

\subsubsection{Global Features}  
Finally, we defined two global features for the graph:  
\begin{itemize}  
    \item $n_A$: The fraction of the system in subsystem `A'.  
    \item $n_B$: The fraction of the system in subsystem `B'.  
\end{itemize}  
The features are summarized in table 1 

\begin{table}[h]
\centering
\caption{Summary of Graph Neural Network Features}
\begin{tabular}{@{}lll@{}}
\toprule
Type & Feature\\ \midrule
Node Features & Coordinates \\
 & Rydberg State Probability\\
 & Subsystem Mask \\ \midrule
Edge Features & Normalized Distance  \\
 & Angle with Horizontal \\
 & Two-Point Correlation \\ \midrule
Global Features & $n_A$  \\
 & $n_B$ \\ \bottomrule
\end{tabular}
\label{tab:features_summary}
\end{table}

We found that this set of features effectively captures the relationships between the atoms and the entanglement entropy, while avoiding unnecessary complexity that could overwhelm the model.

\subsection{Model Architecture}
Our model was constructed using PyTorch Geometric library and consists of the following components:

\subsubsection{Node \& Edge Encoding}
We implemented node and edge encoders as 2-layer multilayer perceptrons (MLPs) with BatchNorm and SiLU (Sigmoid Linear Unit) activation. Each MLP transformed the input features (4-dimensional for nodes, 3-dimensional for edges) to a hidden representation of dimension 512.

\subsubsection{Edge Attention Layers}
We introduced an MLP designed to give the model learnable weights for the importance of separate edges. The edge attention mechanism consisted of a 3-layer MLP applied at each message-passing layer, producing attention weights in the range [0, 1] through a sigmoid activation.

\subsubsection{Edge Representation MLP}
We took the output of the edge encoding MLP as input and generated another edge representation with a MLP consisting of 2 layers, BatchNorm, SiLU activation, and dropout.

\subsubsection{Message Passing Layers}
We implemented a multi-layer architecture for edge-node co-processing, alternating between two distinct message-passing mechanisms across 6 layers. For even-indexed layers, we employed GINEConv\cite{hu2019strategies} operations with enhanced edge features. Each GINEConv utilized a 3-layer MLP that processed node representations, consisting of two linear transformations with BatchNorm, SiLU activation, and dropout ($p=0.4$) between layers. For odd-indexed layers, we leveraged TransformerConv\cite{shi2020masked} operations with edge-aware attention mechanisms. Each TransformerConv implemented multi-head attention with 8 attention heads, where each head processed features of dimension 64. The TransformerConv incorporated beta-transformations and concatenated the outputs from different attention heads. Both convolution types incorporated edge features multiplied by their weights of dimension 512 to guide the message-passing process, enabling rich interactions between node and edge representations throughout the network.

Residual connections were added after each message-passing layer to stabilize training and improve gradient flow. Specifically, the output of each convolution operation $h_{\text{new}}$ was added to the previous node features $h$.

After every even-indexed message-passing layer, we applied an intermediate processing step to further refine the node features. This step consisted of a 2-layer MLP with BatchNorm, SiLU activation, and dropout. The MLP transformed the node features while preserving their dimensionality, allowing for additional feature extraction and regularization.

At the end of each message-passing layer, we updated the edge representation and weights using separate MLPs and used the updated representations as input for the next message-passing layer.

\subsubsection{Graph-Level Readout}
For obtaining a graph-level readout, we implemented a multi-head readout mechanism using two Set2Set modules with 4 processing steps each.

To manage the increased dimensionality from the multi-head readout (2048), we employed a dimension reduction projection. This projection consisted of a linear transformation followed by BatchNorm, SiLU activation, and dropout regularization, reducing the representation to 1024.

\subsubsection{Global Features MLP}
We added an MLP consisting of three layers, BatchNorm, SiLU activation, and dropout to process the global features.

\subsubsection{Final MLP}
The final MLP combined the graph-level readout and the global features MLP output, producing a single scalar value representing the von Neumann entropy. The final MLP consisted of four layers with BatchNorm, SiLU activation, and dropout, followed by a Softplus activation to ensure non-negative output.

A flow chart of the model is shown in figures ~1 and ~2.

\begin{figure}[h]
    \centering
    \includegraphics[width=0.9\linewidth]{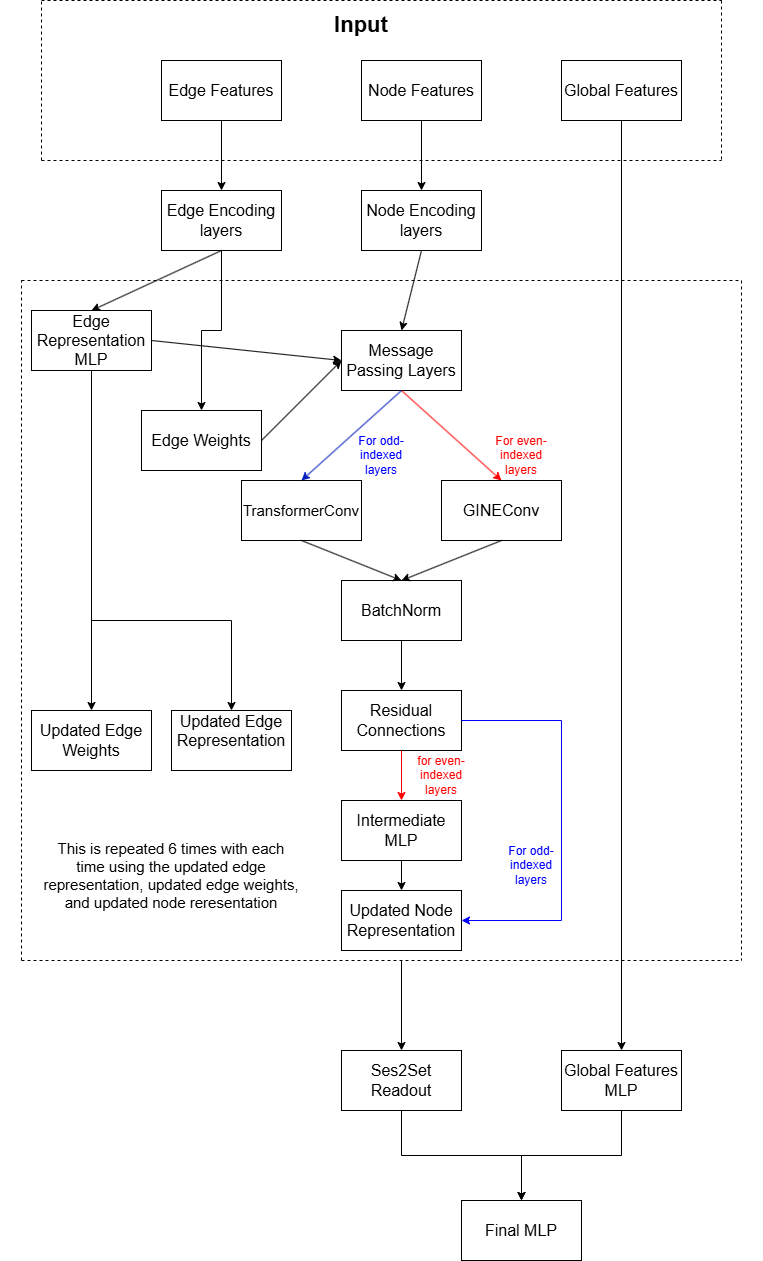}
    \caption{A flow chart of the GNN model, red paths are only taken by even-indexed layers, and blue paths are only taken by odd-indexed layers.}
    \label{fig:flowchart}
\end{figure}

\begin{figure}[h]
    \centering
    \includegraphics[width=0.9\linewidth]{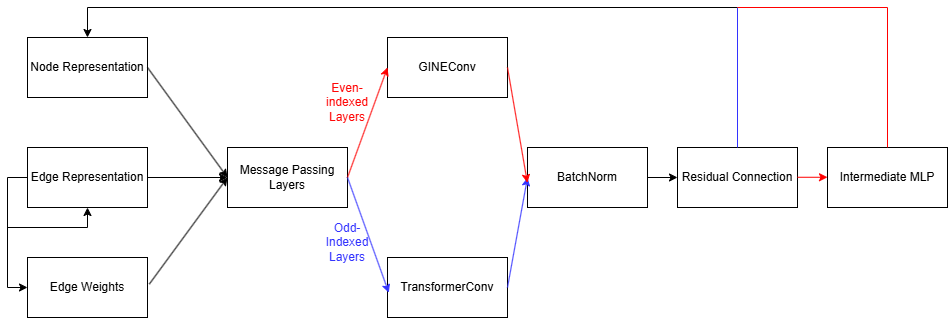}
    \caption{A flow chart of the message passing blocks, red paths are only taken by even-indexed layers, and blue paths are only taken by odd-indexed layers.}
    \label{fig:flowchart}
\end{figure}

\subsection{Training}
We trained the model for 500 epochs using Kaiming initialization\cite{he2015delving}. We used the AdamW optimizer\cite{loshchilov2017decoupled} with a learning rate of $1.5 \times 10^{-4}$ and weight decay of $10^{-4}$. To stabilize the training, we applied gradient clipping with a maximum gradient norm of 1.0. We also used the CosineAnnealingWarmRestarts scheduler\cite{loshchilov2016sgdr} with the following parameters: $T_0 = 50$, $T_{\text{mult}} = 2$, and $\eta_{\text{min}}=10^{-6}$. The exact values for the optimizer parameters, scheduler parameters, dropout rate, hidden layer dimensionality, and the number of message-passing layers were determined using optimization code that minimized our validation loss over 100 epochs. We specifically used a Tree-structured Parzen Estimator\cite{bergstra2011algorithms} over those parameters and obtained the values used in the final training code.

\section{Results}
We used three metrics to evaluate our model. The first was our loss function defined as:
\begin{align}
\text{loss} =\frac{1}{N} \sum \log\left( \cosh(S_{\text{pred}} - S)\right),
\end{align}
the second was the mean absolute error defined as:
\begin{align}
\text{MAE} = \frac{1}{N} \sum \left| S_{\text{pred}} - S \right|,
\end{align}
and the third was the mean absolute percentage error defined as:
\begin{align}
\text{MAPE} = \frac{100}{N} \sum \left| \frac{ S_{\text{pred}} - S }{S}\right|.
\end{align}
However, we only evaluated the MAPE for points that had an actual entropy value higher than 0.01 of the maximum value in the dataset. This threshold approach prevented artificially inflated error metrics in regions of near-zero entropy, where even small absolute errors could result in disproportionately large percentage errors.

After training the model for 500 epochs, the model achieved a validation loss of $1.5\times 10^{-5}$ and a validation MAE of $3.6\times 10^{-3}$. We evaluated the entire dataset after training and recorded the results in Tables~2 and 3, and plotted them in Fig.~3.

\begin{center}
\begin{table}[h]
\caption{Mean absolute error for each system size}
\begin{tabular}{@{}llll@{}}
\toprule
System Size & MAE$\times 10^{-3}$ & System Size & MAE$\times 10^{-3}$ \\ \midrule
2  & 7.386 & 8  & 3.035 \\
4  & 4.439 & 10 & 3.013 \\
6  & 2.801 & 12 & 3.868 \\ \bottomrule
\end{tabular}
\label{tab:mae_results}
\end{table}
\end{center}

\begin{figure}[h]
    \centering
    \includegraphics[width=1\linewidth]{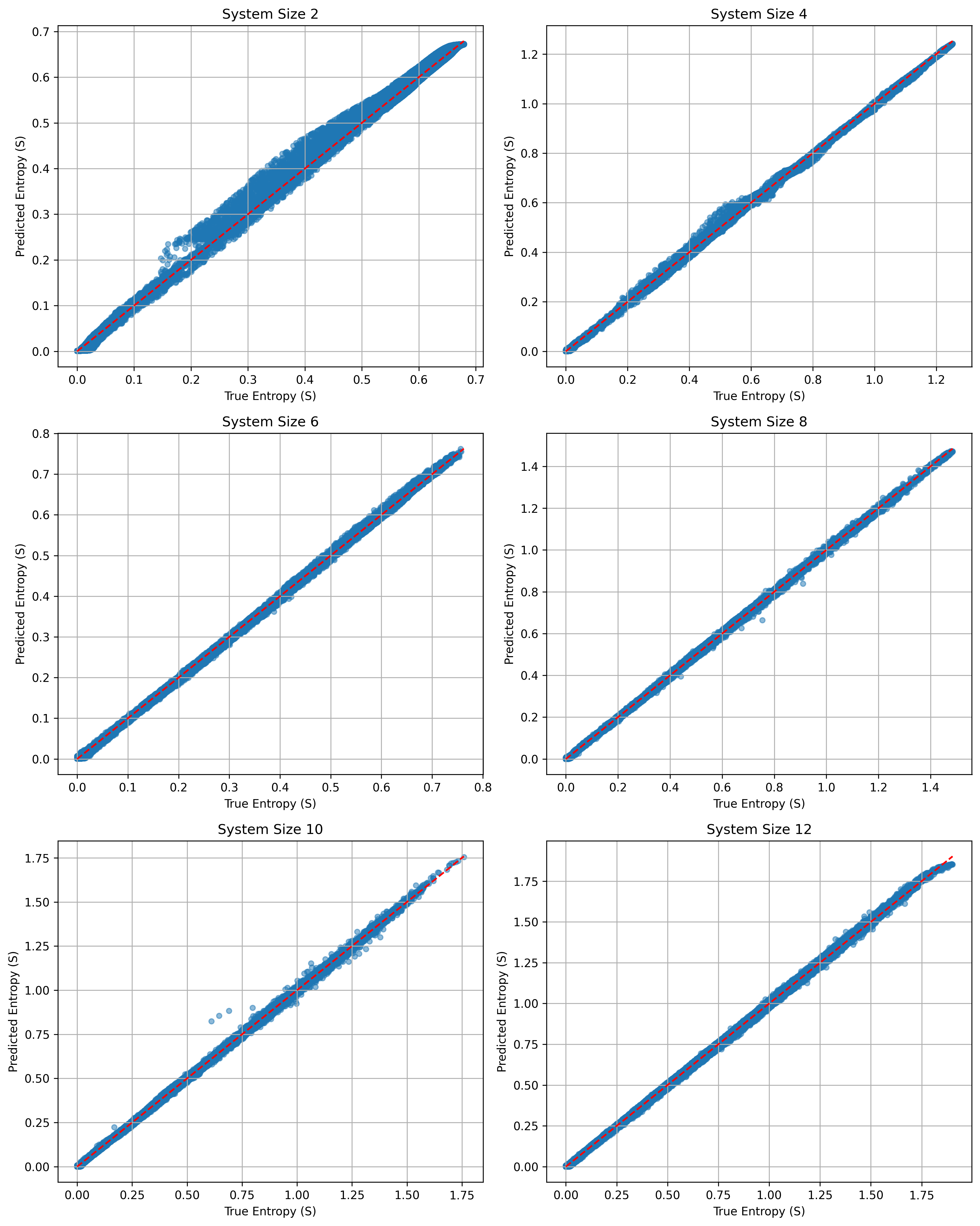}
    \caption{Predicted (y-axis) vs. actual (x-axis) von Neumann entanglement entropy. The red dashed line represents perfect prediction.}
    \label{fig:results}
\end{figure}

\begin{center}
\begin{table}[h]
\caption{Mean absolute percentage error for each system size}
\begin{tabular}{@{}llll@{}}
\toprule
System Size & MAPE & System Size & MAPE \\ \midrule
2  & 4.33\% & 8  & 1.47\% \\
4  & 1.62\% & 10 & 1.67\% \\
6  & 3.89\% & 12 & 1.27\% \\ \bottomrule
\end{tabular}
\label{tab:mape_results}
\end{table}
\end{center}

\subsection{Uncertainty Quantification}

We evaluated the predictive uncertainty of our model using Monte Carlo dropout\cite{gal2016dropout}. We ran our prediction 50 times per point and took the 2.5th and 97.5th percentiles to construct a 95\% confidence interval (CI). However, the original model was overconfident, only giving 83.6\% coverage, so we scaled the spread of our predictions by a temperature parameter $T$ while maintaining the mean of our predictions. We obtained an optimal scaling temperature $T=1.11$ for our 95\% CI, which gives us an average interval width of 0.053. We plotted the coverage vs temperature in Fig. 4, and the model predictions with their calibrated intervals are shown in Fig. 5.

\begin{figure}[h]
    \centering
    \includegraphics[width=1\linewidth]{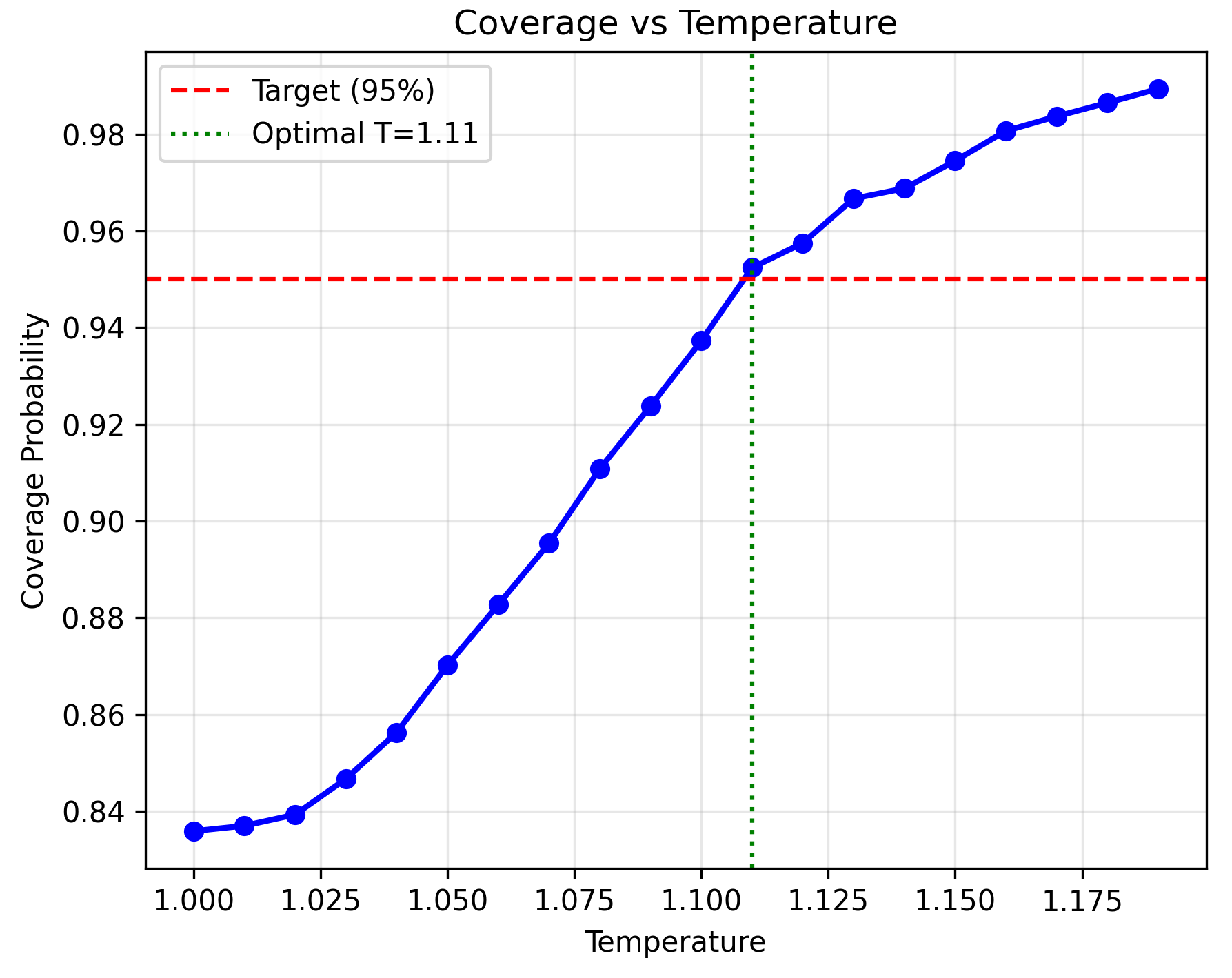}
    \caption{Coverage of the model predictions(y-axis) vs temperature parameter(x-axis)}
    \label{fig:heatmap}
\end{figure}

\begin{figure}[h]
    \centering
    \includegraphics[width=1\linewidth]{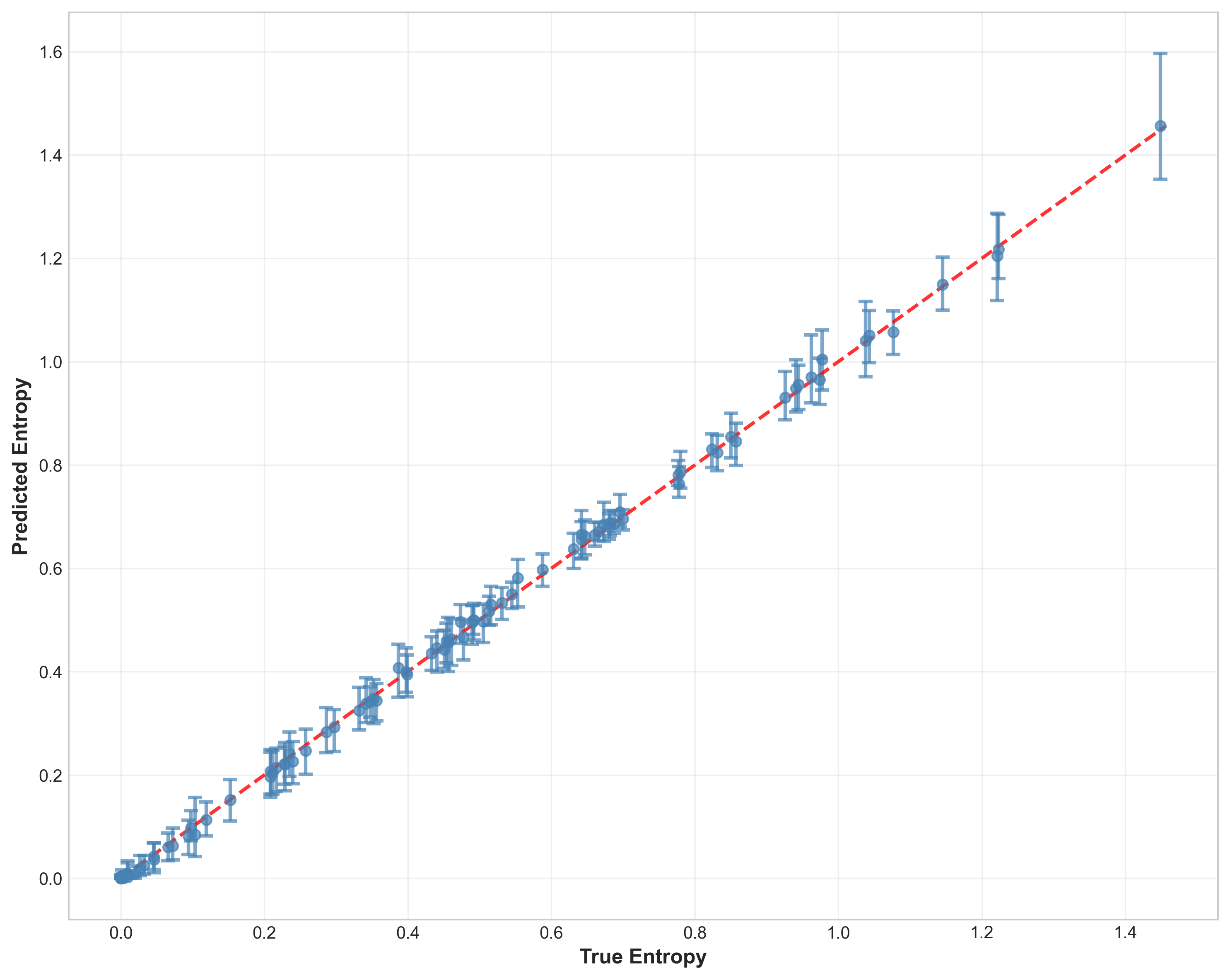}
    \caption{Predicted vs. actual von Neumann entanglement entropy with calibrated 95\% confidence intervals. A representative subset of 100 points is shown for clarity. The red dashed line indicates perfect prediction.}
    \label{fig:heatmap}
\end{figure}

\subsection{Comparative Analysis}
We conducted a paired-t test to compare the accuracy of our model compared to the classical mutual information (MI) estimation on 50000 points. Both methods showed statistically significant systematic bias (GNN: mean bias = 0.0007, t(49999) = 32.51, p $<$ 0.001; MI: mean bias = 0.069, t(49999) = 122.51, p $<$ 0.001), though the GNN bias was substantially smaller.

Comparing the errors of both methods we see that the GNN significantly outperformed the MI approach (t(49999) = 223.73, p $<$ 0.001, Cohen's d = 1.40). The mean absolute error was reduced from 0.103 (MI) to 0.0035 (GNN), representing a 96.6\% improvement. We Plotted the error distributions of both methods in Fig. 6.

\begin{figure}[h]
    \centering
    \includegraphics[width=1\linewidth]{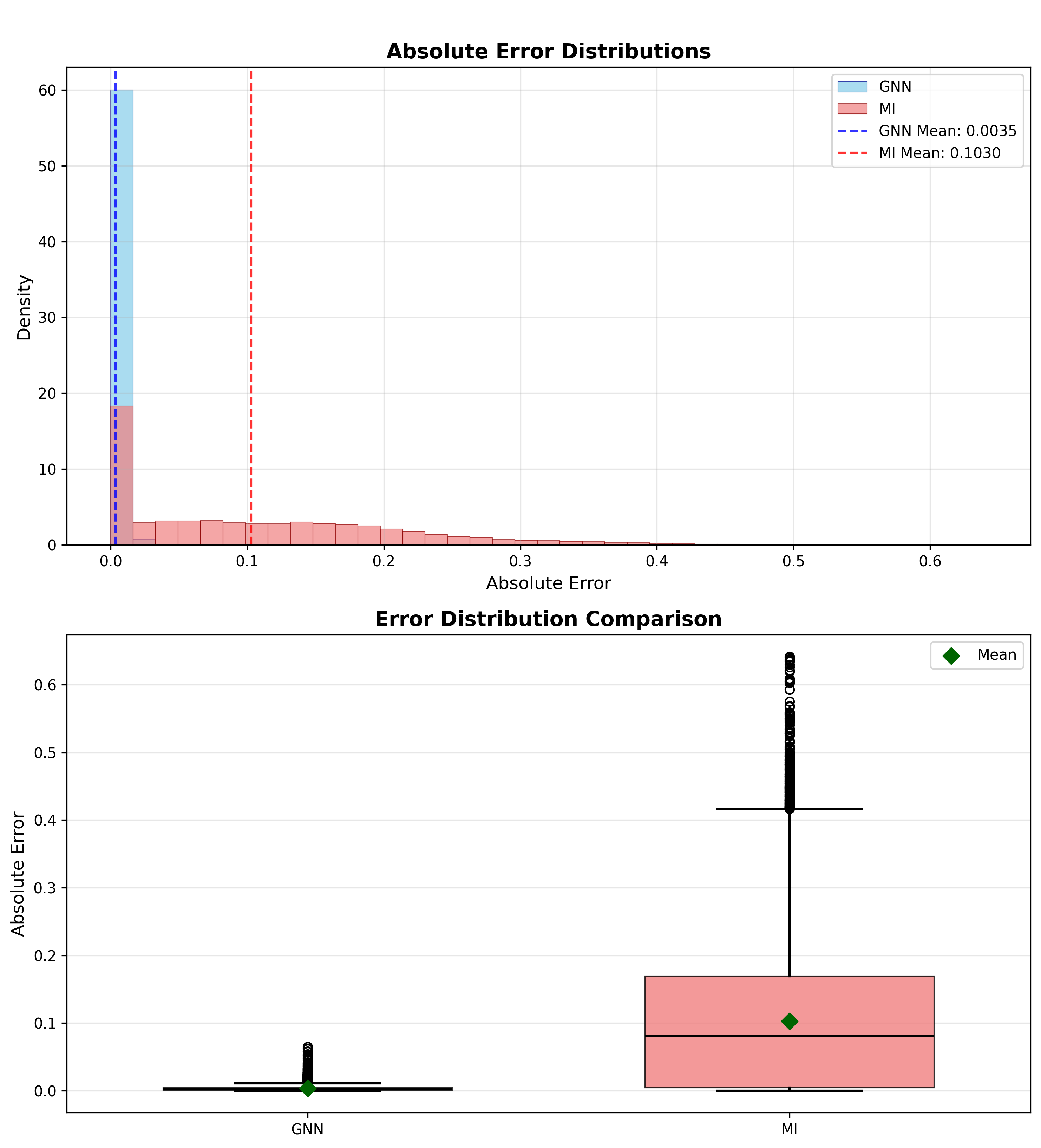}
    \caption{(Top) Probability density distributions of absolute errors for GNN-based (blue) and MI-based (red) Von Neumann entropy estimation. (Bottom) Box plots comparing error distributions of both methods. Both plots were made on a 50000 points sample}
    \label{fig:comparing}
\end{figure}

We also plotted in Fig.~7 the von Neumann entropy, MI, and our model predictions in the phase space of $\frac{R_b}{a}$ and $\frac{\Delta}{\Omega}$ for a symmetrical partition of a 6-rung system, and we can clearly see that our model prediction is much better than the classical MI estimation.

\begin{figure}[h]
    \centering
    \includegraphics[width=1\linewidth]{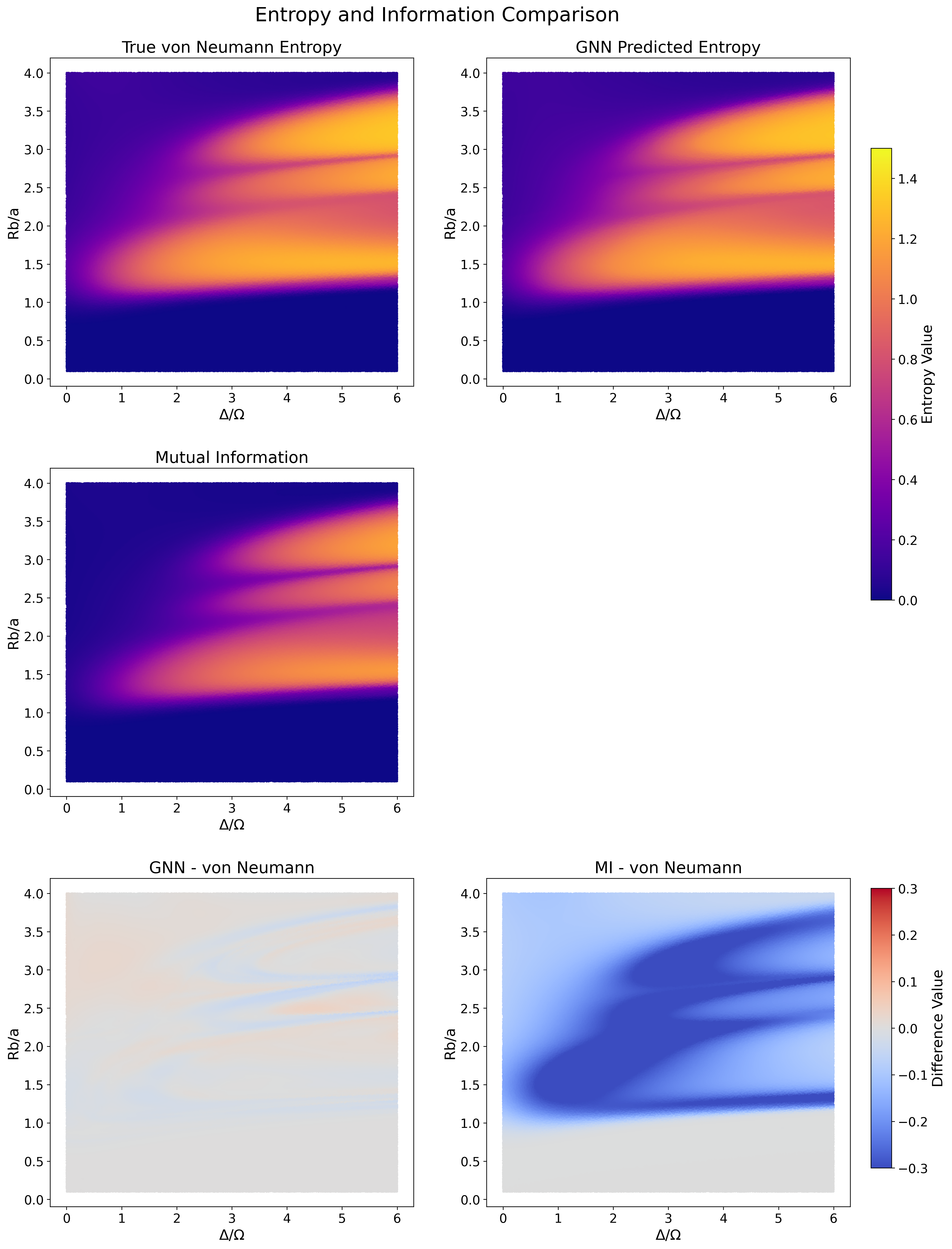}
    \caption{Comparison of von Neumann entropy estimates across parameter space. Heat maps show (top row) true von Neumann entropy and GNN-predicted entropy, (middle) mutual information, and (bottom row) prediction errors (GNN - von Neumann and MI - von Neumann) as functions of $\frac{\Delta}{\Omega}$ and $\frac{R_b}{a}$}
    \label{fig:heatmap}
\end{figure}

\subsection{Model Robustness}
We also tested the model for robustness against small fluctuations in the input, which is very important for any model that depends on experimental data. First we tested the model with limited input data, we accomplished this by instead of using the full probabilities from our matrix diagonalization data, we sample 10000 points from the probability distribution, this effectively simulates an experiment with 10000 readings. We obtained a MAE of 0.007738 and a MAPE of 4.12\% on the symmetric partition dataset, compared to the MAE and MAPE obtained before truncating our wavefunction, which was 0.008720 and 3.84\% respectively. Our results can be seen in Fig.~8.

Another experimental error simulation we did was a bit-flip readout error where each bit has a 1\% chance of flipping, we did this by randomly applying the bit flip on a sampled 10000 states from the probability distribution. We achieved a MAE of 0.016812 and a MAPE of 5.33\%, the results are plotted in Fig. 9.

We also tested the model robustness for wrong classification of the atoms subsystem, we did this by flipping the subsystem of the atoms at the boundary between the two subsystems with a probability of 1\%, the results are shown in figure 10, and all of the experimental errors analysis are summarized in table 4.

\begin{figure}[h]
    \centering
    \includegraphics[width=1\linewidth]{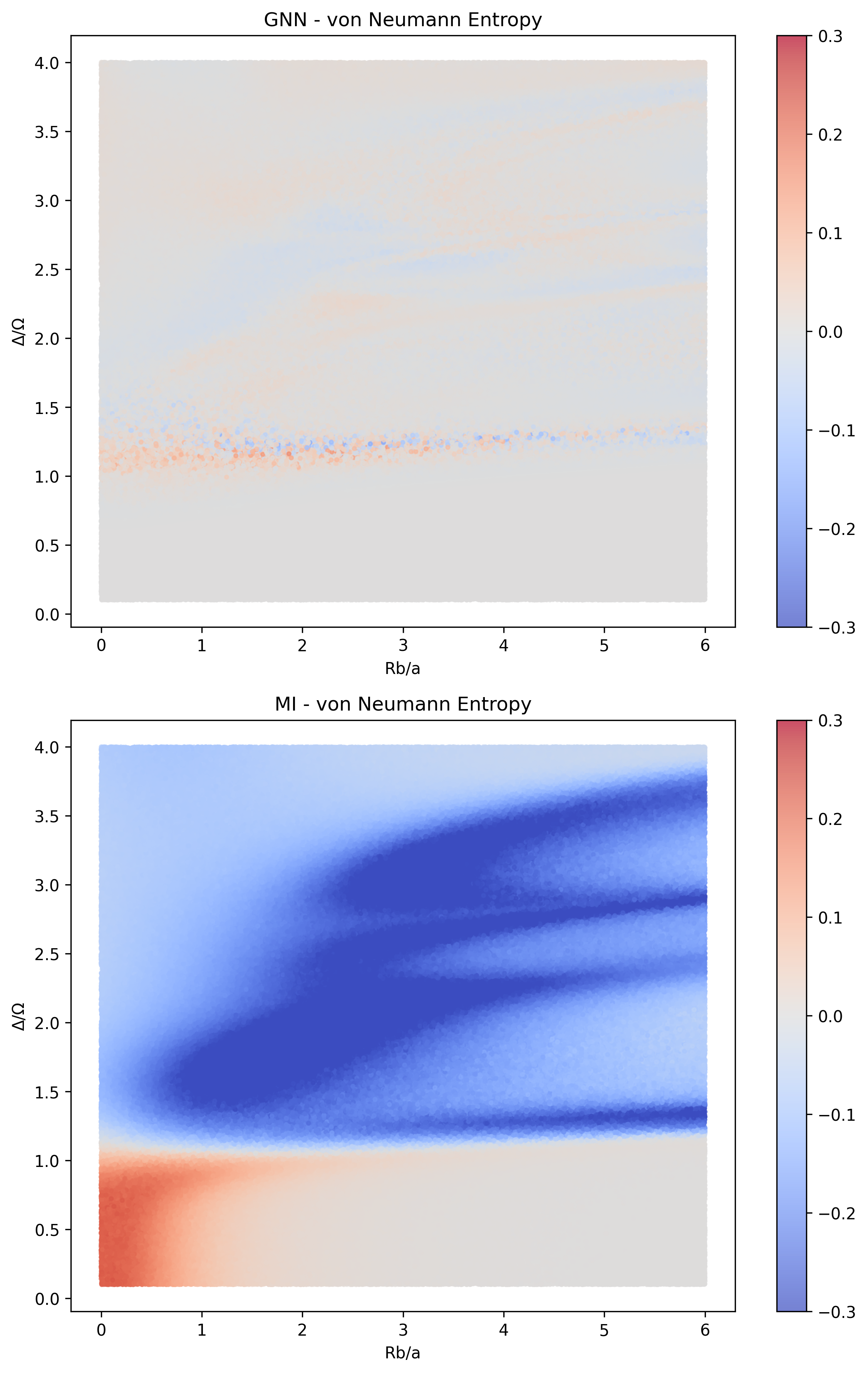}
    \caption{Heat maps of the (Top) GNN - von Neumann entropy, and the (Bottom) MI - von Neumann entropy with limited input data.}
    \label{fig:heatmap_truncated}
\end{figure}

\begin{figure}[h]
    \centering
    \includegraphics[width=1\linewidth]{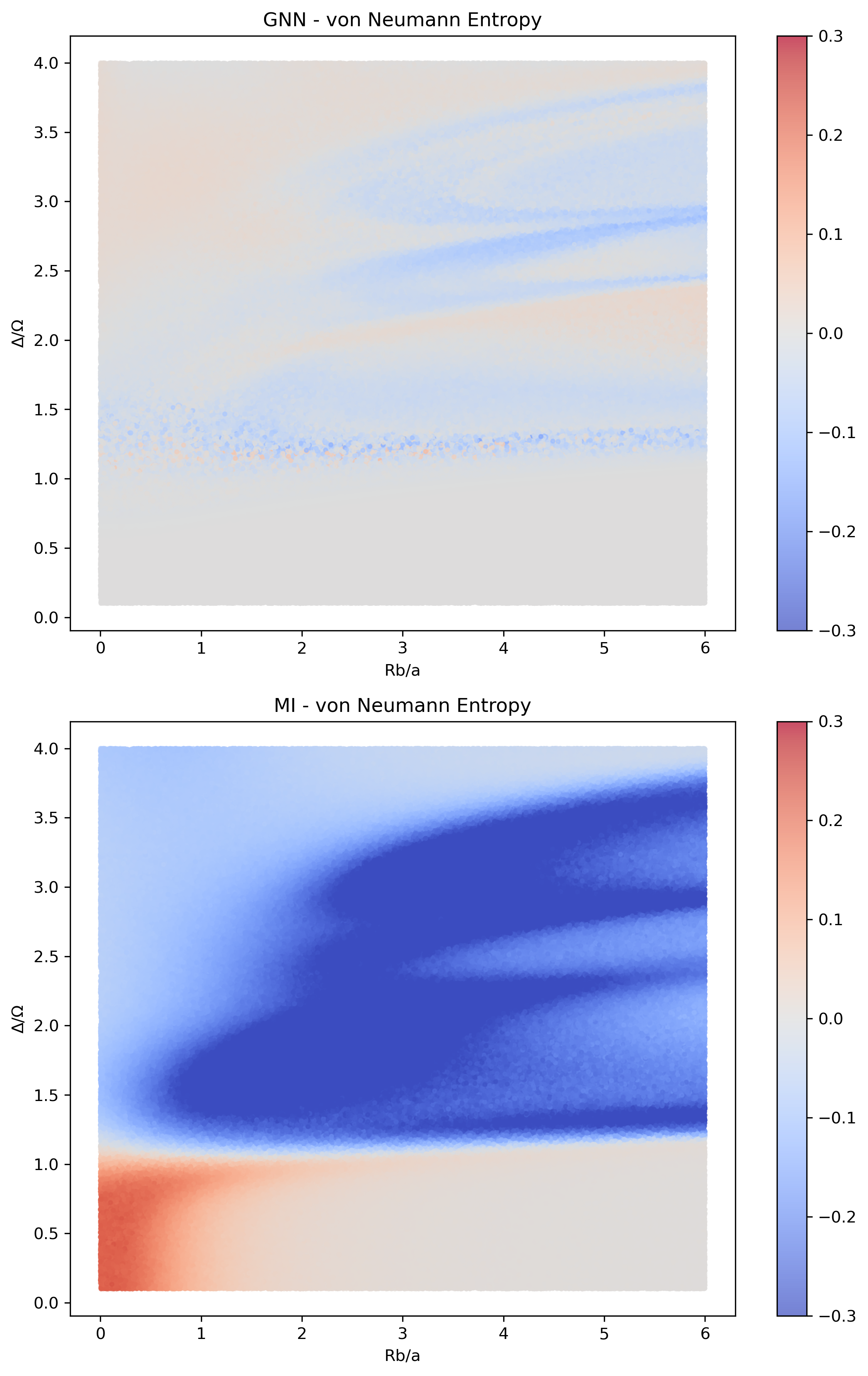}
    \caption{Heat maps of the (Top) GNN - von Neumann entropy, and the (Bottom) MI - von Neumann entropy with limited input data and readout bit-flip errors.}
    \label{fig:heatmap_truncated}
\end{figure}

\begin{figure}[h]
    \centering
    \includegraphics[width=1\linewidth]{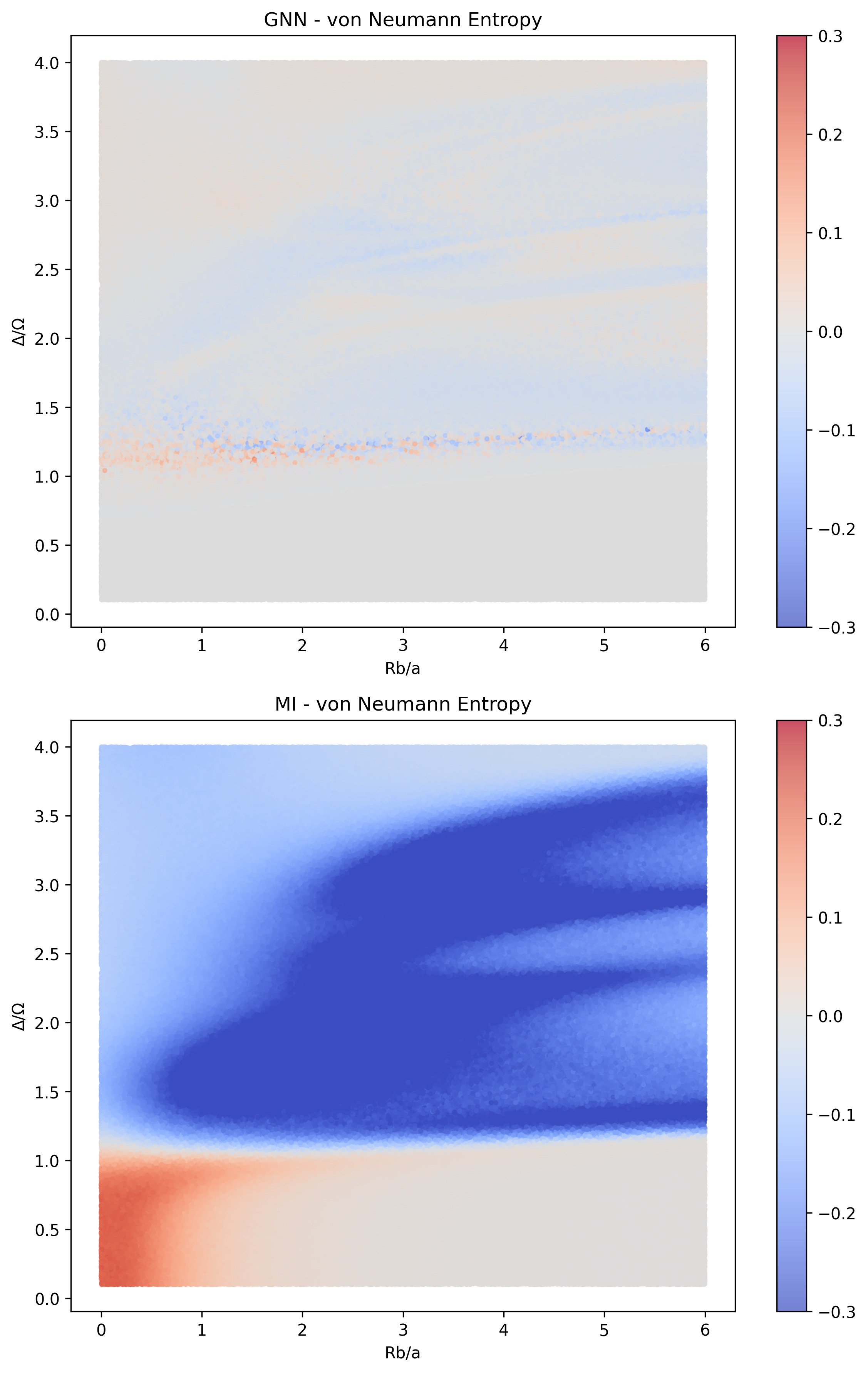}
    \caption{Heat maps of the (Top) GNN - von Neumann entropy, and the (Bottom) MI - von Neumann entropy with limited input data and boundary miss-classification errors.}
    \label{fig:heatmap_truncated}
\end{figure}

\begin{center}
\begin{table}[h]
\caption{Comparing the MAE and MAPE for different experimental errors}
\begin{tabular}{@{}llll@{}}
\toprule
Kind of Error  & MAE$\times 10^{-3}$ & MAPE\\ \midrule
No Experimental Errors  & 8.72 & 3.84\% \\
Limited Sample   & 7.738 & 4.12\% \\
Limited Sample and Readout Errors  & 16.812 & 5.33\% \\
Limited Sample and Classification Errors  & 13.682 & 4.74\% \\
\bottomrule
\end{tabular}
\label{tab:mae_comparison}
\end{table}
\end{center}

Looking also at Cohen's d of the two prediction datasets we can see that it increases when we include experimental errors, signaling that our model might have learned quantum correlations that the classical mutual information doesn't have causing its prediction degradation to be lower, we listed the Cohen's d values in table 5.

\begin{center}
\begin{table}[h]
\caption{Cohen's d between our GNN predictions and the MI for different experimental errors}
\begin{tabular}{@{}llll@{}}
\toprule
Kind of Error  & Cohen's d\\ \midrule
No Experimental Errors  & 1.803\\
Limited Sample   & 1.935 \\
Limited Sample and Bit-Flip Errors  & 1.870\\
Limited Sample and Classification Errors  & 1.911\\
\bottomrule
\end{tabular}
\label{tab:mae_comparison}
\end{table}
\end{center}

\subsection{Fine-Tuning the Model}
Another important feature the model should possess is generalization, as generating data for more than 6 rungs in the same volumes we used to train the model becomes computationally expensive rapidly. Therefore, we tested our model on datasets of 7-10 rungs at specific values for $\frac{\Delta}{\Omega}$. The results are shown in Tables~6 and 7.

While our model outperformed MI for sizes larger than those it was trained on, the MAE and MAPE were not within acceptable error margins. This degradation occurs because the model was trained exclusively on systems with 1-6 rungs, limiting its ability to capture the scaling behavior of larger quantum systems. To address this limitation, we explored fine-tuning as a practical solution for extending the model's applicability to larger systems of interest.

We fine-tuned the model weights for 7 and 8 rungs using a significantly smaller dataset of 10,000 samples, compared to the much larger original training dataset. We trained the model for 200 epochs using a $5\times 10^{-5}$ learning rate and $1.5\times 10^{-4}$ weight decay. We achieved a validation loss of $1.84\times 10^{-4}$ and MAE of $1.0495\times 10^{-2}$, which represented a significant improvement over the original model. We plotted our results for symmetrical partitions in Fig.~11, and the MAE and MAPE are listed in Tables~6 and 7. We also computed the relative improvements the fine tuning provide and listed them in in table 6. 

Fine-tuning achieved substantial improvements across all system sizes, with an overall average improvement of 49.1\%. Notably, the model showed excellent transferability, with 9-10 rung systems (outside the fine-tuning range) achieving lower, but significant improvements (13.9-54.6\%) compared to the 7-8 rung systems (44.3-66.1\%) used for fine-tuning. This demonstrates that the learned representations capture scalable entanglement features that generalize beyond the immediate training scope.

We investigated data augmentation for fine-tuning by generating multiple bipartition configurations per parameter set, effectively multiplying our training samples by a factor of $n$. However, this approach degraded performance, likely because the model overfit to the limited range of $\frac{R_b}{a}$ and $\frac{\Delta}{\Omega}$ values in the small dataset.

\subsection{Targeted Fine-Tuning}
Up to this point, all of our training and fine-tuning was on random partitions and sampled over the entire phase space, while this is great to maintain the generality of the model we also tested how would fine-tuning for a specific partition and part of the phase space would preform, so we took the model trained on 1-6 rungs and fine-tuned it separately on the 4 datasets in figure 11, taking only 100 points for the training, all the data points are partitioned symmetrically and have a $\frac{\Delta}{\Omega} =2.5$, the results are shown in figure 12 and table 8. These results show that fine-tuning for a specific set of parameters could be very effective without requiring much data. 

\begin{center}
\begin{table}[h]
\caption{Performance comparison before and after fine-tuning with improvement metrics}
\begin{tabular}{@{}lllll@{}}
\toprule
Number of & $\frac{\Delta}{\Omega}$ & MAE$\times 10^{-2}$ & MAE$\times 10^{-2}$ & Improvement \\
rungs & & (Before Fine- & (After Fine- & (\%) \\
& & -Tuning)&-Tuning)& \\ \midrule
7  & 2.5 & 3.145 & 1.388 & 55.9\% \\
7  & 3.5 & 3.819 & 1.516 & 60.3\% \\
8  & 2.5 & 6.555 & 2.223 & 66.1\% \\
8  & 3.5 & 5.528 & 3.081 & 44.3\% \\
9  & 2.5 & 6.936 & 3.834 & 44.7\% \\
9  & 3.5 & 4.2231 & 3.637 & 13.9\% \\
10 & 2.5 & 14.977 & 6.801 & 54.6\% \\
10 & 3.5 & 12.747 & 5.967 & 53.2\% \\
\bottomrule
\end{tabular}
\label{tab:mae_comparison}
\end{table}
\end{center}

\begin{center}
\begin{table}[h]
\caption{Comparing the MAPE for the original and fine-tuned model}
\begin{tabular}{@{}llll@{}}
\toprule
Number of & $\frac{\Delta}{\Omega}$ & MAPE & MAPE \\
rungs & & (Before Fine-Tuning) & (After Fine-Tuning) \\ \midrule
7  & 2.5 & 15.22\% & 12.92\% \\
7  & 3.5 & 13.92\% & 8.61\% \\
8  & 2.5 & 15.83\% & 6.99\% \\
8  & 3.5 & 11.85\% & 6.82\% \\
9  & 2.5 & 22.21\% & 12.73\% \\
9  & 3.5 & 10.66\% & 8.44\% \\
10 & 2.5 & 30.26\% & 14.35\%  \\
10 & 3.5 & 19.28\% & 8.71\% \\
\bottomrule
\end{tabular}
\label{tab:mape_comparison}
\end{table}
\end{center}

\begin{figure}[h]
    \centering
    \includegraphics[width=1\linewidth]{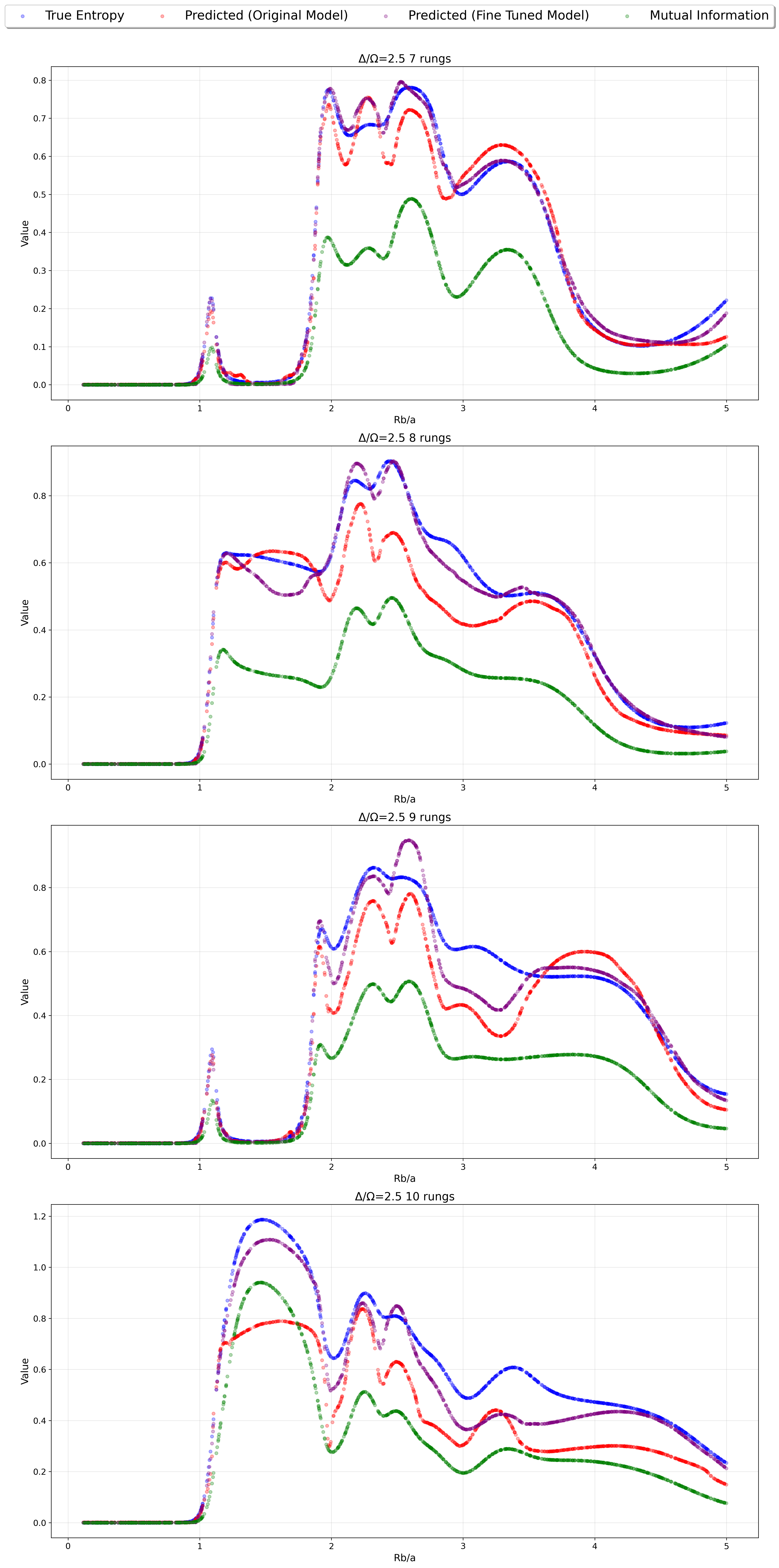}
    \caption{Von Neumann entropy (blue), the original model (red), the fine-tuned model (purple), and the MI (green) as a function of $\frac{R_b}{a}$.}
    \label{fig:comparison_grid}
\end{figure}

\begin{figure}[h]
    \centering
    \includegraphics[width=1\linewidth]{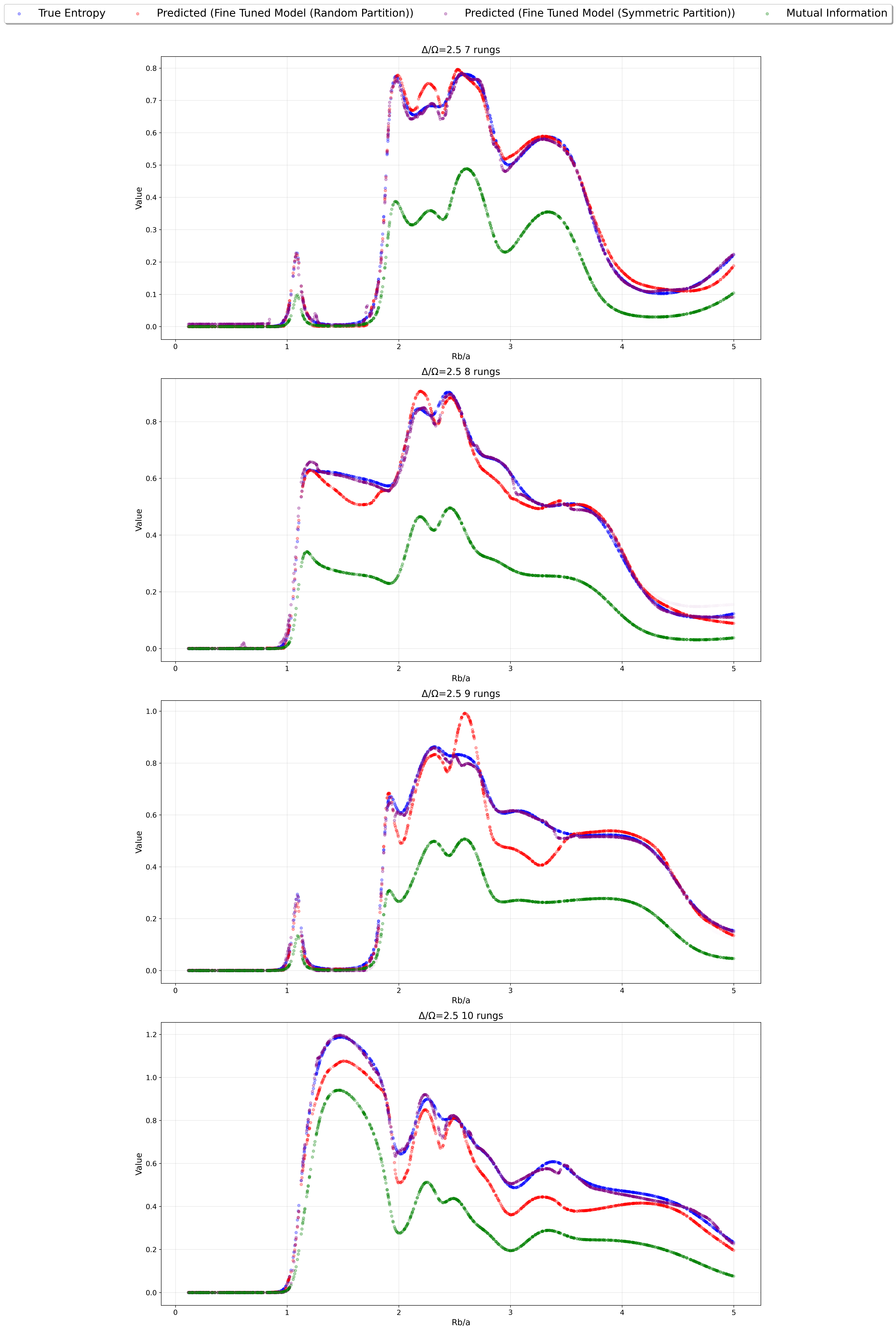}
    \caption{Von Neumann entropy (blue), the model fine-tuned on 10000 points of 7 and 8 rungs randomly partitioned(red), the model fine-tuned on 100 point of a symmetric partition of the specific rung with $\frac{\Delta}{\Omega} =2.5$(purple), and the MI (green) as a function of $\frac{R_b}{a}$.}
    \label{fig:comparison_grid}
\end{figure}

\begin{center}
\begin{table}[h]
\caption{Performance comparison of 2 different fine-tuning approaches discussed in sections D and E}
\begin{tabular}{@{}lllll@{}}
\toprule
Number of & MAE$\times 10^{-2}$ & MAE$\times 10^{-2}$ & Improvement \\
rungs  & (Fine- & (Targeted Fine- & (\%) \\
& -Tuning)&-Tuning)& \\ \midrule
7  & 1.388 & 0.769 & 44.6\% \\
8   & 2.223 & 0.879 & 60.5\% \\
9 & 3.834 & 0.775 & 79.8\% \\
10  & 6.801 & 1.282 & 81.1\% \\
\bottomrule
\end{tabular}
\label{tab:mae_comparison}
\end{table}
\end{center}

\section{Limitations}

This study has several limitations that warrant further investigation. First, while our model demonstrated reasonable generalization performance, the significant increase in MAE for larger systems and the substantial improvement from fine-tuning suggests that the model learned entropy representations but failed to fully capture system size scaling laws. Future studies training on datasets with broader size distributions may better teach the model proper scaling behavior.

Second, the model's physical interpretation requires deeper analysis. Comparing results with DMRG calculations and examining the hidden layer representations could provide valuable insights into the underlying physics,

Third, while fine-tuning significantly improved performance on larger system sizes (7–10 rungs), this approach assumes consistent entanglement scaling behavior across system sizes. If the underlying physics changes significantly with scale, fine-tuning with small datasets may fail to generalize, potentially requiring more comprehensive retraining with larger datasets.

\section{Discussion \& Conclusion}
In this paper, we proposed an architecture for a graph neural network model that processes experimentally accessible data and predicts the von Neumann entanglement entropy. We tested this model on a Rydberg system similar to the one used in the Aquila quantum computer developed by QuEra.

Our model demonstrated excellent performance within its training range, achieving a mean absolute error of $3.6\times 10^{-3}$ and a mean average percentage error of 1.44\%. When tested outside its training range, the model still outperformed the classical mutual information approximation, though with noticeably reduced accuracy. This limitation could be addressed by fine-tuning the model for specific system sizes of interest using small datasets. We demonstrated this approach by starting with our base model and fine-tuning it on 7- and 8-rung systems. We observed significant improvements in our predictions for 7-10 rung systems.

Furthermore, we showed that when focusing on a small region of the phase space and a specific partitioning scheme, we could achieve a mean absolute error comparable to our original training performance ($\sim 10^{-3}$) using very small fine-tuning datasets (100 points in our examples). This approach could be extended to any number of rungs by generating small datasets using appropriate approximation methods.

\section*{Code Availability}
The codes used to generate the dataset, preprocess it, train the model, and fine-tune it, along with the weights of the original and fine-tuned model, are available on the GitHub repository: \url{https://github.com/AsalehPhys/Predicting-von-Neumann-Entropy}

\section*{Acknowledgments}
We would like to thank Professor Yannick Meurice for the valuable discussions that shed light on the specific system we studied and its properties. We would also like to thank Leen Saleh for her guidance in exploring the possible machine learning models that could work for our study.

\appendix

\section{Architecture Design Choices}
We evaluated several alternative GNN architectures and plotted the models MAE as a function of training epoch. First we Tested a pure GINEConv model and a pure TransformerConv model in opposition to our hybrid approach, the results can be seen in figure 13.

\begin{figure}[h]
    \centering
    \includegraphics[width=1\linewidth]{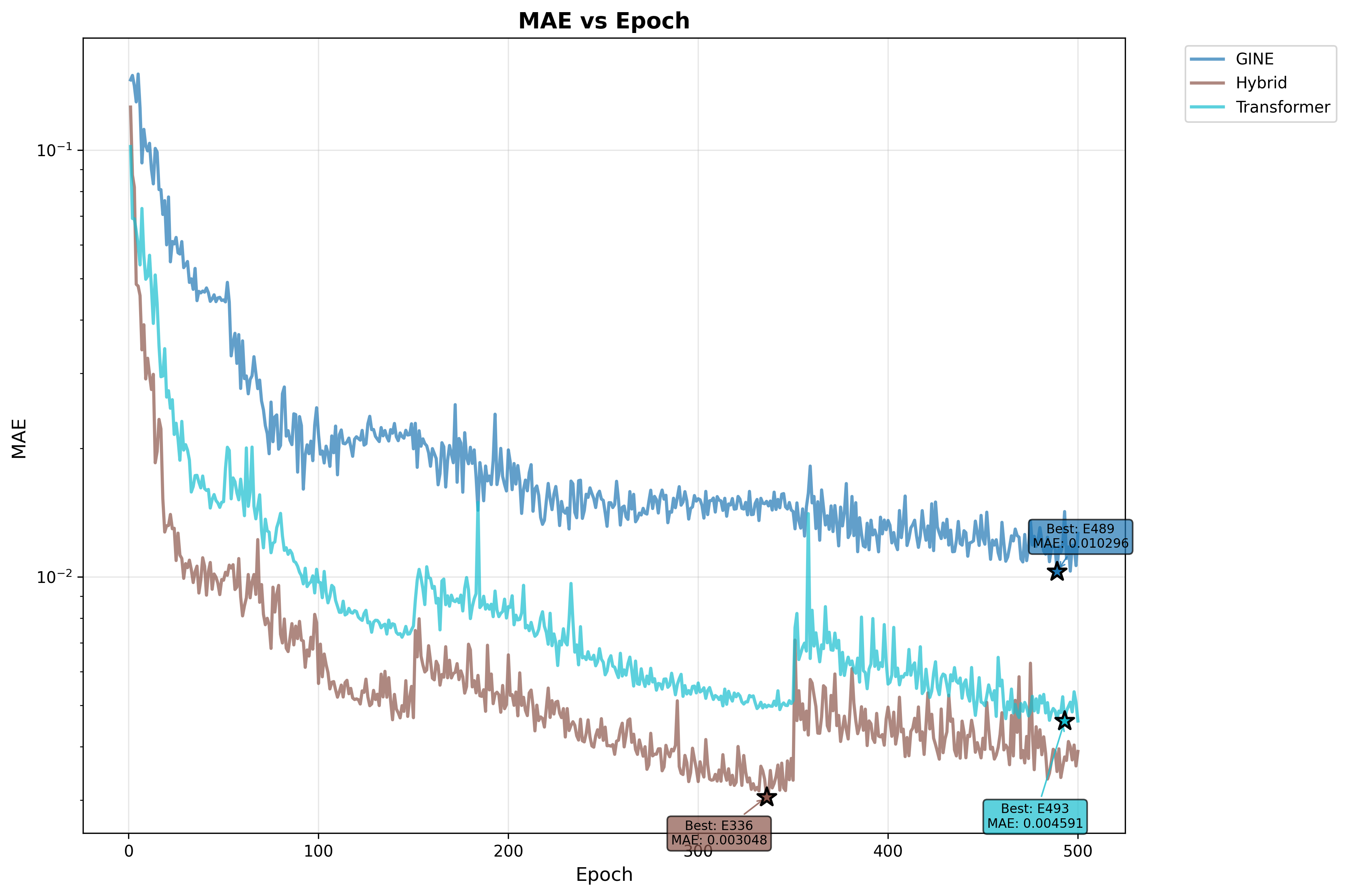}
    \caption{Validation set MAE as a function of training epochs for a pure GINEConv model, a pure TransformerConv model and our hybrid approach}
    \label{fig:comparison_grid}
\end{figure}

We also evaluated different readout mechanisms and plotted our results in figure 14.

\begin{figure}[h]
    \centering
    \includegraphics[width=1\linewidth]{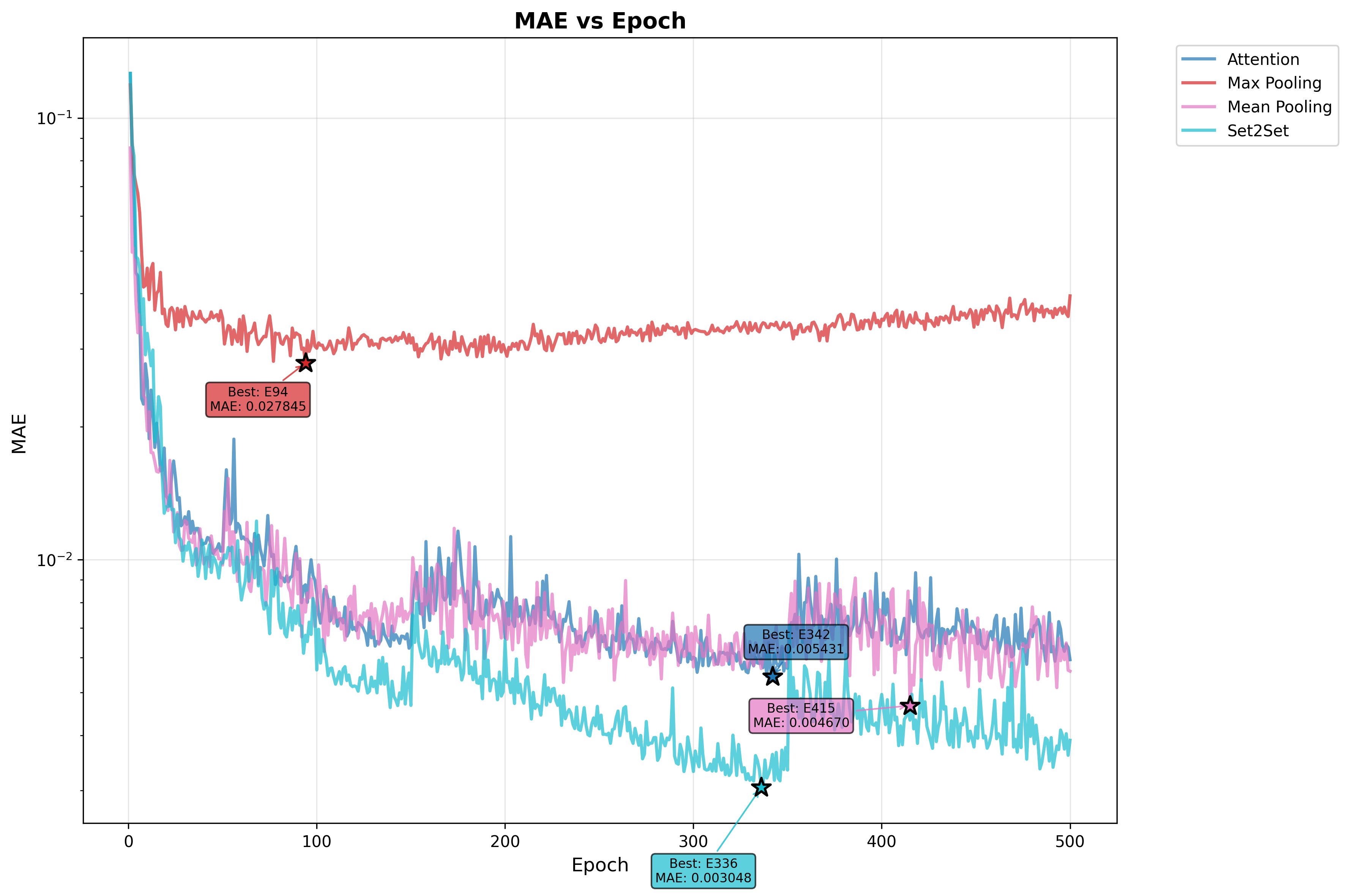}
    \caption{Validation set MAE as a function of training epochs for a max pooling, mean pooling, set2set, and attention readout}
    \label{fig:comparison_grid}
\end{figure}

Looking at the edge features, we tested the inclusion of higher order moments to see if that improve the information propagation through our model, we tested the forth cross-moment $E[(x_i-p_i)^2 (x_j-p_j)^2]$ where the $x$ is a binary value representing if the site is excited and $p$ is the excitation probability. We didn't find any improvements, the results are shown in figure 15.

\begin{figure}[h]
    \centering
    \includegraphics[width=1\linewidth]{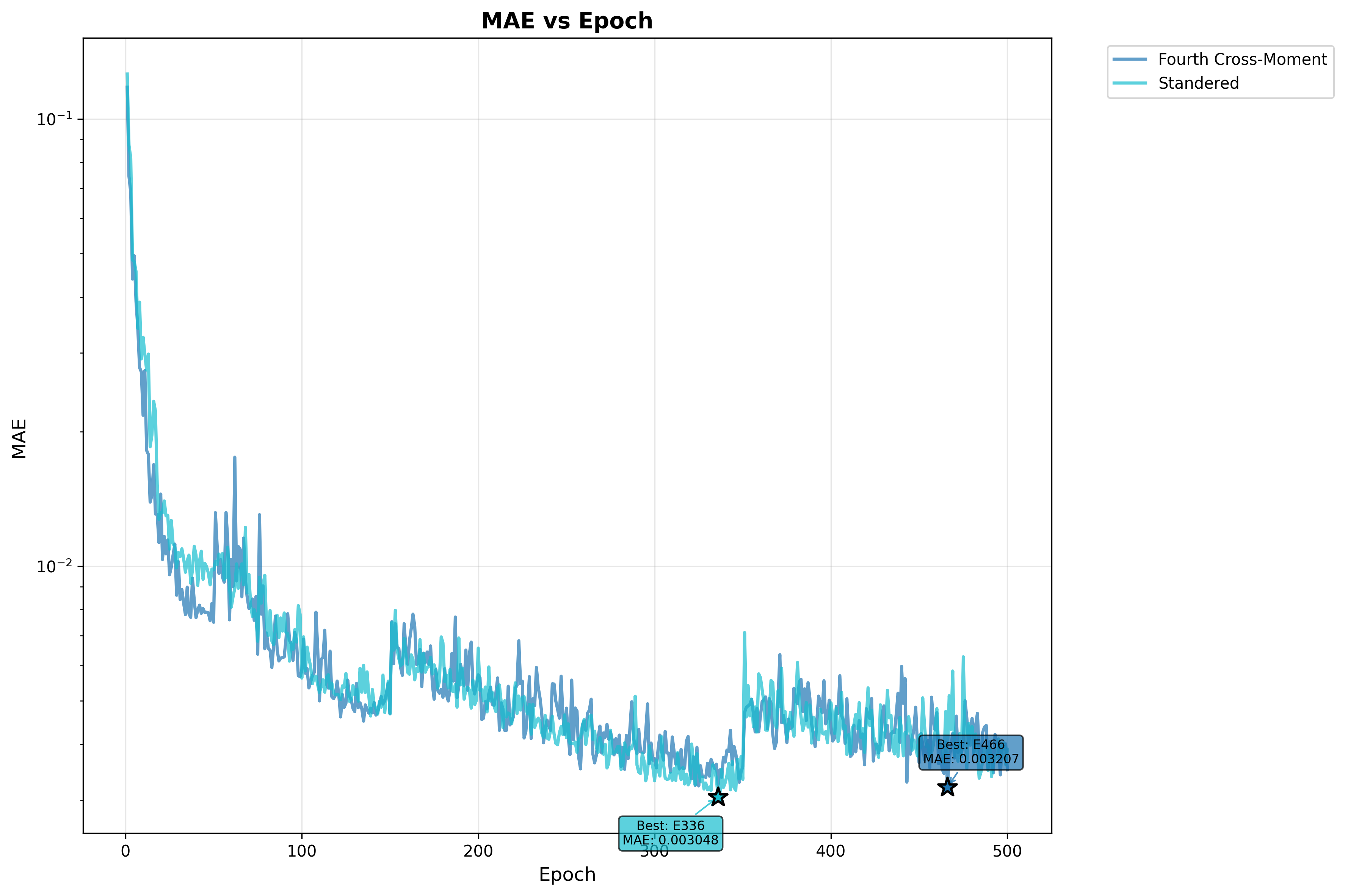}
    \caption{Validation set MAE as a function of training epochs for 2 model with the forth cross-moment added as an edge feature to one of them}
    \label{fig:comparison_grid}
\end{figure}

Lastly, we evaluated how well does the model preform if we had a distance cutoff for graph edges instead of creating fully connected graphs. Our results can be shown in figure 16, while we can clearly see improvements for choosing a cutoff for edge creation, this makes the model generalization worse especially when trying to predict far larger systems than what it was trained on. This is likely due to the inherit non-locality of the entanglement entropy. We show in figure 17 the generalization of 2 models, one that was trained on fully connected graphs and the other was trained on graphs with an edge creation cutoff of 4 units.

\begin{figure}[h]
    \centering
    \includegraphics[width=1\linewidth]{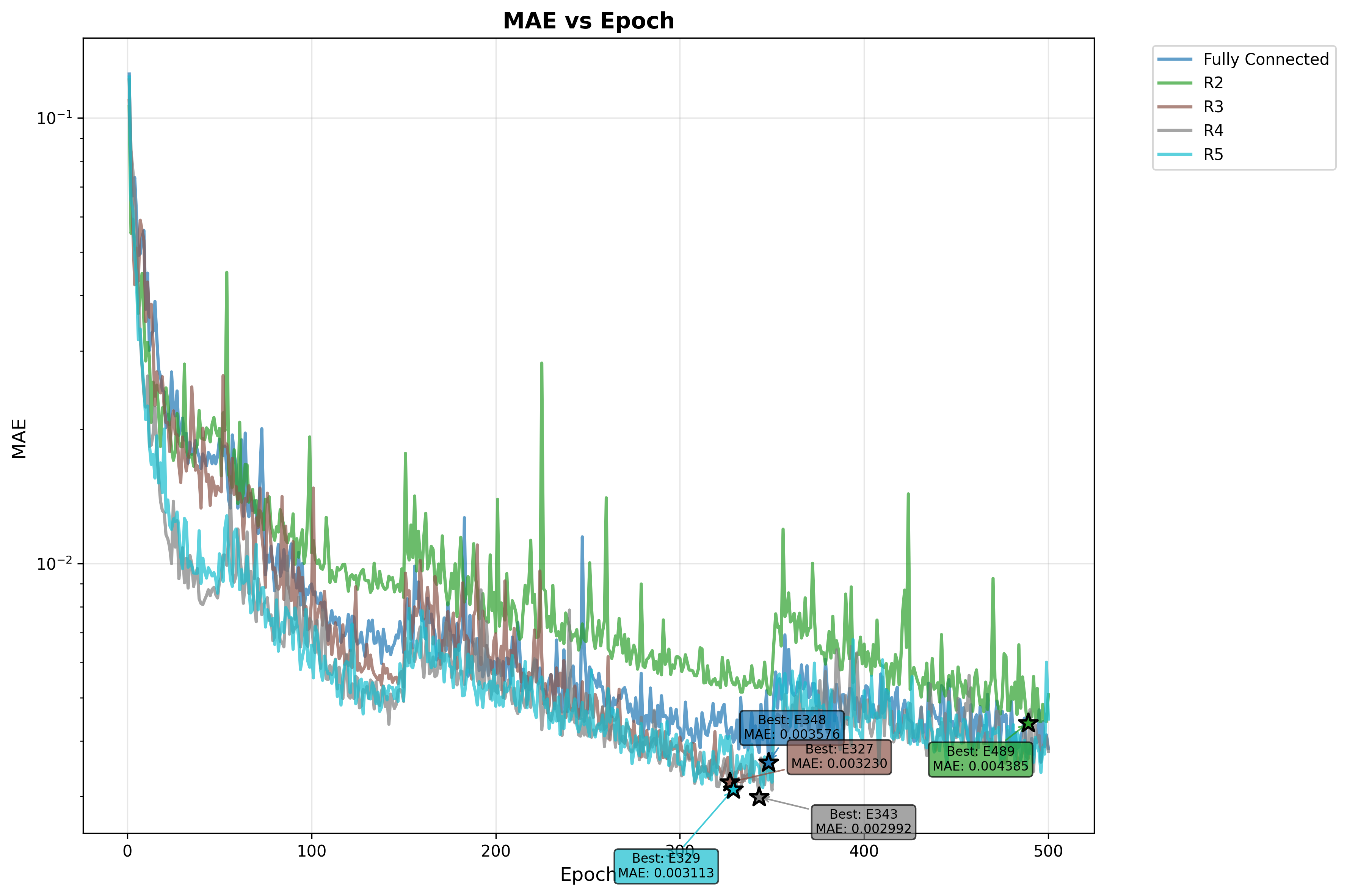}
    \caption{Validation set MAE as a function of training epochs for different values of edge creation cutoff distance from 2 units to fully connected}
    \label{fig:comparison_grid}
\end{figure}

\begin{figure}[h]
    \centering
    \includegraphics[width=1\linewidth]{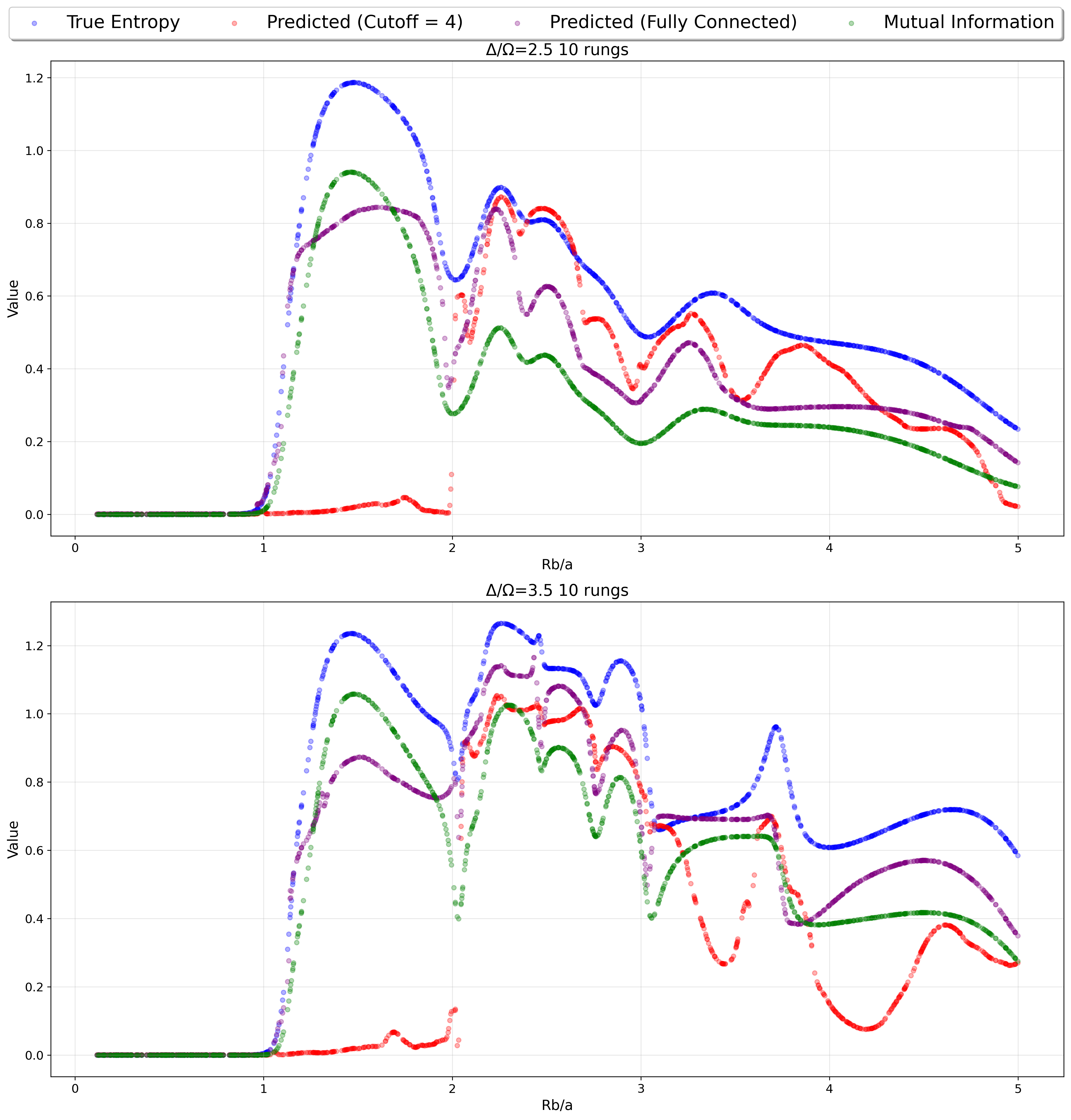}
    \caption{Comparision of the genralization of 2 models, one that was trained on partially connected graphs(red) and one that was trained on fully connected graphs(purple)}
    \label{fig:comparison_grid}
\end{figure}

\newpage

\bibliographystyle{apsrev4-2} 
\bibliography{references}

\end{document}